\documentclass{article}
\usepackage{latexsym}
\usepackage{color,cite,graphicx,subfigure,setspace}% to put in axodraw
\usepackage{latexsym,amssymb,epsf} % don't remember if these are
\newcommand{\bfc}{{\bf c}}
\newcommand{\bfE}{{\bf E}}
\newcommand{\bfJ}{{\bf J}}
\newcommand{\bfn}{{\bf n}}
\newcommand{\bfp}{{\bf p}}
\newcommand{\bfP}{{\bf P}}
\newcommand{\bfx}{{\bf x}}

\newcommand{\bfsigma}{{\mbox{\boldmath $\sigma$}}}

\newcommand{\calE}{{\cal E}}
\newcommand{\calJ}{{\cal J}}
\newcommand{\calL}{{\cal L}}
\newcommand{\calN}{{\cal N}}
\newcommand{\calP}{{\cal P}}
\newcommand{\calZ}{{\cal Z}}
\newcommand{\bfxt}{\left(\bfx,t\right)}
\newcommand{\bfxtp}{\left(\bfx+\bfc_i,t\right)}
\begin{document}
\title{
   Three dimensional hydrodynamic lattice-gas simulations of binary immiscible and ternary amphiphilic flow through porous media.
}
\author{
  Peter J. Love\\
  {\footnotesize Theoretical Physics, Department of Physics, University of Oxford,}\\
  {\footnotesize 1 Keble Road, Oxford, OX1 3NP, UK}\\
  {\footnotesize{\tt love@thphys.ox.ac.uk}}\\[0.3cm]
  Jean-Bernard Maillet,\\
  {\footnotesize CEA, 31/33 rue de la F{\'e}d{\'e}ration}\\
  {\footnotesize 75752 Paris}\\
  {\footnotesize{\tt maillet@bruyeres.cea.fr}}\\
  and\\[0.3cm]
  Peter V. Coveney,\\
  {\footnotesize Centre for Computational Science,
    Queen Mary,}\\
  {\footnotesize University of London, Mile End Road,}\\
  {\footnotesize London E1 4NS, U.K.}\\
  {\footnotesize{\tt p.v.coveney@qmw.ac.uk}}\\
}
\maketitle

\begin{abstract}
We report the results of a study of multiphase flow in porous media. A Darcy's law for steady multiphase flow was investigated for both binary and ternary amphiphilic flow. Linear flux-forcing relationships satisfying Onsager
reciprocity were shown to be a good approximation of the simulation
data. The dependence of the relative permeability coefficients on water saturation was investigated and showed good qualitative agreement with experimental data. Non-steady state invasion flows were investigated, with
particular interest in the asymptotic residual oil saturation. The
addition of surfactant to the invasive fluid was shown to
significantly reduce the residual oil saturation. 
\end{abstract}

\section{Introduction}

The study of fluid flow in porous media is of great industrial importance. Applications include enhanced oil recovery, aquifer purification, containment of toxic and nuclear waste, geological flows of magma, chemical reactions in catalysts, and the study of blood flow through capillaries. The fluids of interest are frequently `multi-functional', for example, certain oil field fluids may be required not only to displace oil and gas and transport them to the surface, but also to act as coolants and lubricants for the drill bit and for the transport of cuttings from within the well. Blood transports salts and sugars in solution, as well as white and red blood cells in colloidal suspension. Such fluids clearly require a more sophisticated description than that provided by single phase continuum approaches.  Subjects of theoretical interest include the relationship between the macroscopic transport coefficients of porous media and their microscopic geometry~\cite{bib:roth3}, the geometrical properties of interfaces~\cite{bib:feder1, bib:feder2, bib:feder3,bib:feder4} and the
relationship between the transport coefficients and the morphology of
the flow~\cite{bib:gunroth1,bib:olroth2}.

The flow of multiphase fluids in arbitrary geometries remains a
challenge for conventional modelling techniques~\cite{bib:anderson,bib:whitaker}.  In a top-down computational fluid dynamics (CFD) approach one would use a finite-difference or similar method to numerically integrate the Navier-Stokes equations with appropriate boundary conditions. Such methods become enormously complex when one wishes to simulate flow in an arbitrary geometry, or when one wishes to introduce the complex
interfaces present in multiphase flow. The alternative atomistic approaches are not viable, as a `typical' rock pore size is on the micron scale, far larger than the nanometre domain accessible with present day molecular dynamics. 

In order to make progress in the study of fluid flow in porous media,
it is therefore necessary to simplify either the rock geometry or the fluid
flow. Reducing the dimensionality of the flow enables larger systems
to be studied computationally. The results from these simulations can
then be compared with experimental results from micromodels~\cite{bib:kad}. In
three dimensions, one can construct a simplified model of a single
pore, and model the medium as a {\it network} of such simplified
pores. The flow of a multiphase fluid through such a network is modelled based on some simplifying assumptions about the flow in each pore. Such models have been used extensively to study steady state flows~\cite{bib:goodramak}. A variation on the network model, `invasion percolation', has been used to investigate the development of interfacial fronts in the invasive flows of interest in oil field situations~\cite{bib:feder1, bib:feder2, bib:feder3,bib:feder4}.

In recent years X-ray micro-tomography techniques have developed sufficient resolution to provide accurate three-dimensional digitizations of real rock structures. It is evidently preferable to study flow processes taking place in such realistic porous media. In order to do this a simplified, computationally efficient model of multiphase flow is required. Mesoscale fluid models present clear advantages over conventional CFD techniques for this application.

Lattice-gas and lattice-Boltzmann models have been used to study flow through porous media in both two and three dimensions. In two dimensions, lattice-gas models have been used to study both binary immiscible and ternary amphiphilic flow~\cite{bib:roth2,bib:roth4,bib:pvcjb,bib:pvcjb2}. In three dimensions the Rothman-Keller model has been used to study binary immiscible flow through a digitized slab of Fontainebleau sandstone~\cite{bib:olroth2}. Studies of binary immiscible flow in three dimensions were also performed using a lattice-Boltzmann model~\cite{bib:gunrothzalzan,bib:gunroth1,bib:gunroth2}.

In this paper the three dimensional amphiphilic lattice gas model, described in detail in~\cite{bib:bcp} is applied to the study of flow through porous media. The adaptations of the model to handle arbitrarily complicated boundaries and fluid forcing are described. An investigation of steady state flows and comparison with Darcy's law is then presented for both two-phase binary immiscible and three-phase ternary amphiphilic flow. The simulation results are discussed in the context of linear nonequilibrium thermodynamics and recent extensions of Onsager theory to multiphase flow in porous media~\cite{bib:onsager1,bib:onsager2,bib:prideflekkoy,bib:flekkoypride,bib:degrootmazur}. The behaviour of non-steady-state, `invasion' flow is then simulated. Such simulations are intended to reproduce oil field flows, and the effect of surfactant on flow morphology and oil extraction is investigated. This work represents an extension of previous results obtained with the two-dimensional lattice gas model~\cite{bib:pvcjb,bib:pvcjb2}.

\section{The lattice gas model}\label{sec:model}

Our lattice-gas model is based on a microscopic, bottom-up
approach, where dipolar amphipile particles are included alongside
the immiscible oil and water species.  Lattice-gas particles can have
velocities $\bfc_i$, where $1\leq i\leq b$, and $b$ is the number of
velocities per site.  We shall measure discrete time in units of one
lattice timestep, so that a particle emerging from a collision at site
$\bfx$ and time $t$ with velocity $\bfc_i$ will advect to site
$\bfx+\bfc_i$ where it may undergo the next collision.
We let $n^\alpha_i\bfxt\in\{0,1\}$ denote the presence ($1$) or absence
($0$) of a particle of species $\alpha\in\{R,B,A\}$ ($R$, $B$, $A$
denoting red (oil), blue (water) and green (amphiphile) species
respectively) with velocity $\bfc_i$, at lattice site $\bfx\in\calL$
and time step $t$. The collection of all $n^\alpha_i\bfxt$ for $1\leq
i\leq b$ will be called the {\it population state} of the site; it is
denoted by
\begin{equation}
\bfn\bfxt\in\calN
\end{equation}
where we have introduced the notation $\calN$ for the (finite) set of
all distinct population states.
The amphiphile particles also have an orientation denoted by
$\bfsigma_i\bfxt$.  This orientation vector, which has fixed magnitude
$\sigma$, specifies the orientation of the amphiphile particle at site $\bfx$ and time step $t$
with velocity $\bfc_i$. The collection of the $b$ vectors
$\bfsigma_i\bfxt$ at a given site $\bfx$ and time step $t$ is called
the {\it orientation state}. 
We also introduce the {\it colour charge} associated
with a given site,
\begin{equation}
q_i\bfxt\equiv n^R_i\bfxt-n^B_i\bfxt,
\end{equation}
as well as the total colour charge at a site,
\begin{equation}
q\bfxt\equiv\sum_{i=1}^b q_i\bfxt.
\end{equation}
The state of the model at site ${\bf x}$ and time
$t$ is completely specified by the
population state and orientation state of all the sites. 
The time evolution of the system is an alternation between an advection
or {\it propagation} step and a {\it collision step}.  In the first of
these, the particles move in the direction of their velocity vectors to
new lattice sites.  This is described mathematically by the replacements
\begin{equation}
n^\alpha_i\left(\bfx+\bfc_i,t+1\right)
\leftarrow
n^\alpha_i\bfxt,
\label{eq:propp}
\end{equation}
\begin{equation}
\bfsigma_i\left(\bfx+\bfc_i,t+1\right)
\leftarrow
\bfsigma_i\bfxt,
\label{eq:propo}
\end{equation}
for all $\bfx\in\calL$, $1\leq i\leq b$ and $\alpha\in\{ R,B,A\}$.  That
is, particles with velocity $\bfc_i$ simply move from point $\bfx$ to
point $\bfx+\bfc_i$ in one time step.
In the collision step, the newly arrived particles interact, resulting
in new momenta and surfactant orientations.  The collisional change in
the state at a lattice site $\bfx$ is required to conserve the mass of
each species present
\begin{equation}
\rho^\alpha\bfxt\equiv\sum_i^b n^\alpha_i\bfxt,
\end{equation}
as well as the $D$-dimensional momentum vector
\begin{equation}
\bfp\bfxt\equiv\sum_\alpha\sum_i^b\bfc_i n^\alpha_i\bfxt,
\end{equation}
(where we have assumed for simplicity that the particles all carry unit
mass).  Thus, the set $\calN$ of population states at each site is
partitioned into {\it equivalence classes} of population states having
the same values of these conserved quantities. 

We assume that the characteristic time for collisional and
orientational relaxation is sufficiently fast in comparison to that of
the propagation that we can model this probability density as the
Gibbsian equilibrium corresponding to a local, sitewise Hamiltonian function; that is
\begin{equation}
\calP(s')
=
\frac{1}{\calZ}\exp\left[-\beta H(s')\right], \label{eq:beta_defn}
\end{equation}
where $\beta$ is an inverse temperature, $H(s')$ is the energy
associated with collision outcome $s'$, and $\calZ$ is the
equivalence-class partition function.
The sitewise Hamiltonian function for  our model has been previously derived and
described in detail for the two-dimensional version of the
model~\cite{bib:bce}, and we use the same one here. It is
\begin{equation}
H(s') =
\bfJ\cdot (\alpha\bfE+\mu\bfP) +
\bfsigma'\cdot (\epsilon\bfE+\zeta\bfP) +
\calJ : (\epsilon\calE+\zeta\calP)+{\delta \over 2}
{{{\bf v}({\bf x},t)}^{2}},
\label{eq:hamil}
\end{equation}
where we have introduced the {\it colour flux} vector of an outgoing
state,
\begin{equation}
\bfJ\bfxt\equiv\sum_{i=1}^b\bfc_i q'_i\bfxt,
\label{eq:cflux}
\end{equation}
the {\it total director} of a site,
\begin{equation}
\bfsigma\bfxt\equiv\sum_{i=1}^b\bfsigma_i\bfxt.
\end{equation}
the {\it dipolar flux} tensor of an outgoing state,
\begin{equation}
\calJ\bfxt\equiv\sum_{i=1}^b\bfc_i\bfsigma'_i\bfxt,
\end{equation}
the {\it colour field} vector,
\begin{equation}
\bfE\bfxt\equiv\sum_{i=1}^b\bfc_i q\bfxtp,
\label{eq:bfEdef}
\end{equation}
the {\it dipolar field} vector,
\begin{equation}
\bfP\bfxt\equiv-\sum_{i=1}^b\bfc_i S\bfxtp,
\end{equation}
the {\it colour field gradient} tensor,
\begin{equation}
\calE\bfxt\equiv\sum_{i=1}^b\bfc_i\bfE\bfxtp,
\end{equation}
the {\it dipolar field gradient} tensor,
\begin{equation}
\calP\bfxt\equiv-\sum_{i=1}^b\bfc_i\bfc_i S\bfxtp,
\label{eq:calPdef}
\end{equation}
defined in terms of the scalar director field,
\begin{equation}
S\bfxt\equiv\sum_{i=1}^b\bfc_i\cdot\bfsigma_i\bfxt.
\label{eq:sdef}
\end{equation}
and the kinetic energy of the particles at a site,
\begin{equation}
{\delta \over 2}\left|{\bf v}({\bf x},t)\right|^2,
\end{equation}
where ${\bf v}$ is the average velocity of all particles at a site, the
mass of the particles is taken as unity, and $\alpha$, $\mu$,
$\epsilon$, $\zeta$ and $\delta$ are coupling constants. To maintain
consistency with previous work we use the coupling constants as previously
defined in  ~\cite{bib:bcp}. The values of these constants
are:
\begin{center}
$\alpha=1.0$,  $\epsilon=2.0$,   $\mu=0.75$,   $\zeta=0.5 $
\end{center}
These values were chosen in order to maximise the desired behaviour of
sending surfactant to oil-water interfaces while retaining the
necessary micellar binary water-surfactant phases. 
It should be noted that the inverse temperature-like
parameter $\beta$ (whose numerical value is varied in this paper) is not related in the conventional way to the kinetic
energy. For a discussion of the introduction of this parameter into
lattice gases we refer the reader to the
original work by Chen, Chen, Doolen and Lee~\cite{bib:2chendoolen}, and Chan and Liang~\cite{bib:chli}.
Eqs.~(\ref{eq:hamil})-(\ref{eq:calPdef}) were derived by assuming that
there is an interaction potential between colour charges, and that the
surfactant particles are like ``colour dipoles'' in this context~\cite{bib:bce}.
The term parameterised by $\alpha$ models the interaction of colour charges
with surrounding colour charges as in the original Rothman-Keller
model~\cite{bib:rk}; that parameterised by $\mu$ describes the interaction of
colour charges with surrounding colour dipoles; that parameterised by
$\epsilon$ accounts for the interaction of colour dipoles with surrounding
colour charges (alignment of surfactant molecules across oil-water
interfaces); and finally that parameterised by $\zeta$ describes the interaction of colour
dipoles with surrounding colour dipoles (corresponding to interfacial bending energy or
``stiffness''). This model has
been extensively studied in two dimensions
~\cite{bib:bce,bib:em1,bib:em2,bib:em3,bib:em4}, and the three-dimensional implementation employed in the present paper is described in
more detail by Boghosian, Coveney and Love~\cite{bib:bcp}.

\subsection{Porous media}

Porous media are introduced into lattice-gas simulations by adding another bit to the description of the state at a lattice site. This bit takes the value one at rock sites and zero at pore states. In order to realistically simulate oil field multiphase flows it is necessary to include the wetting properties of the porous media. Within the lattice-gas model described above, the rock sites are assigned a colour charge $w$ which may range from $-26 \leq w \leq 26$. Thus $w=-26$ represents a rock which is completely water wetting. This implementation of rock sites is identical to that used in the two-dimensional implementation of the model~\cite{bib:pvcjb,bib:pvcjb2}.

The porous medium used for all ensuing simulations is a subset of a Fontainebleau sandstone X-ray microtomography dataset. The original dataset was obtained by Exxon Mobil research using the synchrotron source at Brookhaven. The total dataset consisted of approximately 19.5 million voxels at a scale of $7.5 \mu m$, over a volume roughly $2$mm on a side. This dataset was also used by Olson and Rothman~\cite{bib:olroth2}. The porosity (ratio of void space to total volume) of the medium is $0.58$. Fontainebleau sandstones are widely used as they are one of the cleanest sandstones available, consisting mainly of reasonably monodisperse quartz sand grains. They also contain low concentrations of paramagnetic iron salts, minimising the difficulties associated with flow imaging by techniques such as nuclear magnetic resonance. The system size chosen for all the following simulations was $64^3$. This is not the largest system size attainable, but was chosen as a reasonable size to allow large numbers of simulations to be performed as we are interested in the behaviour of our model over a wide range of fluid composition and fluid forcing levels, as well as requiring considerable ensemble averaging to obtain reasonable precision on our measurements.

\subsection{Fluid forcing}

Fluid forcing is implemented in the code in order to reproduce the effect of one fluid experiencing a buoyancy effect due to gravity. Forcing is carried out by selecting random sites across the lattice and adding or removing momentum as required. Momentum is added until a fixed amount of momentum per lattice site of the forced fluid has been added. The forcing level referred to below is the average amount of momentum added per forced fluid site. Forcing in is in the positive $z$-direction.

\subsection{Fluid-solid boundary conditions}

Two types of boundary conditions can be used at fluid-solid boundaries: slip and no-slip conditions. In the slip condition, a particle incident on the solid wall undergoes specular reflection, that is, the particle preserves its component of momentum parallel to the wall, and has its component of momentum perpendicular to the wall reversed. In no-slip boundary conditions, particles incident on the wall have their momentum reversed, leading to the alternative description of `bounce-back' conditions. In CFD simulations the no-slip condition is typically used, in that case by imposing a zero tangential fluid velocity at the boundary.

No-slip boundary conditions may be trivially implemented for lattice-gas and lattice-Boltzmann simulations. Particles (or single particle distribution functions in the case of lattice-Boltzmann) which advect onto a site occupied by a rock particle are moved into states with the opposite momentum. This will correspond to a zero velocity condition between the lattice sites adjacent to the boundary and the boundary itself~\cite{bib:rz}. However, if the fluid has a wetting interaction with the solid, such a boundary condition may not be appropriate. Recent non-equilibrium molecular dynamics simulations have illustrated this, showing that a significant slip velocity may exist for such situations~\cite{bib:mdwet}. Mixed boundary conditions allowing for a non-zero slip velocity at the wall could be introduced into the lattice-gas model by including more complex fluid-solid collision rules. However, this possibility remains unexplored, and we do not pursue it in the following simulations.

\subsection{Simulation box boundary conditions}

A typical porous medium is not periodic, and so periodic boundary conditions at the edge of the simulation box cannot be used. 
It is important to keep some periodicity in the direction of the flow to maintain continuity of the flow of fluids. In order to achieve periodicity in the direction of the flow, a mirror image of the porous medium is built in the direction of the flow.
The use of the mirror image ensures that all the pores are connected,
and so periodic boundary conditions can be used in this
direction. This method was first used by Olson and Rothman in their three dimensional lattice gas studies~\cite{bib:olroth,bib:olroth2}. For the boundary conditions in the directions transverse to the flow it is not practical to use the same method because it increases the box size by a factor of two for each direction. 

There are two possible boundary conditions that can be applied: `wall conditions' or `buffer conditions'. The wall conditions consist of creating a plane of obstacle sites at the simulation box edge in the transverse directions. Every particle coming to the wall is normally bounced back. The buffer conditions consist of removing a
region of width two lattice sites around the faces of the box and then placing a
wall around the faces. Particles can then flow in and out of this
reservoir into the pore space. The different boundary conditions were tested on a simulation of a one component fluid flowing in a $32^3$ porous medium. A reference value for the permeability has been chosen using a $64^3$ porous medium and calculating the flow only in the core (assuming that the boundary conditions do not influence the flow in the core of the material). Using the wall
conditions the permeability is half the reference permeability, while using the buffer conditions, a value close to the reference value was found. The buffer conditions have therefore been used for all subsequent simulations. 

For invasion simulations the boundary conditions in the flow direction must simulate the effect of the system being connected to an infinite reservoir of invasive fluid at the `bottom' of the system. At the `top' of the system the boundary conditions must reproduce the effect of a reservoir of fluid displaced from the porous media. For the binary case where the invasive fluid is water penetrating a medium saturated by oil the boundary conditions are relatively simple. Oil particles leaving the top of the system are transferred to the bottom and recoloured, so that an equal number of water particles enter the bottom of the system. The total number of particles in the simulation is conserved, although the composition of the system obviously changes. 

For the ternary invasion case the situation is more complex. The boundary conditions described above create an artificial oil-water interface at the system boundaries in the flow direction. The surfactant particles show a strong tendency to adsorb at these `interfaces'. This gives rise to two simulation artefacts. Firstly, adding a constant amount of surfactant to the system at each timestep does not give a water:surfactant ratio in the invasive fluid equal to that added. This is because the surfactant particles adsorb to the fictitious oil-water interface at the bottom of the system rather than entering the porous medium. The second spurious effect occurs when the invasive fluid front reaches the top of the system. The `oil reservoir' simulated by the upper boundary condition is then in contact with a high concentration water-surfactant solution. This water surfactant solution then solubilises oil out of the oil reservoir above the system, leading to an increase in the oil concentration in the system. Clearly both these effects are highly undesirable.

The first artefact can be removed by actively controlling the amount of surfactant injected to compensate for the loss of surfactant to the `interface' at the bottom of the system. The second artefact is also simply removed: one replaces the oil reservoir by a reservoir with the same composition as the top layer of sites in the system. 

\section{Steady-state flow - Darcy's law}

We validate an extension of the single phase Darcy's law to multiphase
flow which explicitly admits a viscous coupling between fluid phases.
The single phase Darcy's law relationship is:
\begin{equation}\label{eq:spdarcy}
{\bf J} = -\frac{k}{\mu}(\nabla p - \rho {\bf g}),
\end{equation}
where ${\bf J}$ is the fluid flux, $k$ is the permeability, $\mu$ is the
kinematic viscosity, and $(\nabla p - \rho {\bf g})$ is the force on
the fluid due to pressure gradients and gravity. 

As previously noted there is no widely accepted hydrodynamic description of multiphase fluids. Similarly there is no universally agreed extension of Darcy's law for multiphase flow. The single phase Darcy relation has been commonly extended as:
\begin{equation}\label{eq:mpdarcy}
{\bf J}_i = \sum_j k_{ij}(S)\frac{k}{\mu_{i}}{\bf X}_j,
\end{equation}
where $k_{ij}(S)$ is the relative permeability coefficient (depending only on
the saturation), ${\bf J_{i}}$ is the flux of the $i$th species and
${\bf X_{i}}$ is the body force acting on the $j$th
component. 

A number of questions arise concerning these phenomenological descriptions. Firstly, what is the domain of validity of the linear relationship between flux and forcing? As the flow is still governed by the (non-linear) Navier-Stokes equations the linear dependence in~(\ref{eq:spdarcy}) and~(\ref{eq:mpdarcy}) must break down at some forcing value. Secondly, the dependence of the permeability coefficients on the rock microgeometry is of considerable interest. Ideally one would like to determine the coefficients $k$ and $k_{ij}$ independently, and then use eqns~(\ref{eq:spdarcy}) and~(\ref{eq:mpdarcy}) to predict flow rates. Finally any symmetry properties of the matrix $k_{ij}$ are of considerable interest from a theoretical standpoint.

Equation~(\ref{eq:mpdarcy}) has the form of a linear phenomenological law. Such descriptions of transport processes form the theory of linear nonequilibrium thermodynamics. The general theory of such processes was first described by Onsager in 1931~\cite{bib:onsager1,bib:onsager2}. On the basis of the reversibility of the microscopic dynamics of atomic motion Onsager derived the reciprocity relations~\footnote{It is interesting to note that in his papers Onsager was unwilling to commit himself as to whether the dynamics of atomic motion was actually irreversible. Writing only six years after Heisenberg's first paper on matrix mechnics, quantum mechanics was insufficiently well established as a reversible theory of atomic dynamics to settle this issue.}
\begin{equation}\label{eq:recip}
k_{12}=k_{21}.
\end{equation}
The validity of these relations additionally requires the thermodynamic forces ${\bf X}_j$ and fluxes ${\bf J}_i$ to be conjugate in the sense that the entropy production can be written $\sum_i {\bf J}_i \cdot {\bf X}_i$.

Flekk{\o}y and Pride~\cite{bib:flekkoypride} recently demonstrated that the fluxes and forces in Darcy's law are indeed reciprocal. By assuming a particular form for the mechanism of forcing they were able to demonstrate the validity of reciprocity in the infinite capillary number limit, where the interface between the fluids does not move. They went on to consider two additional cases, firstly where the interface relaxes to its equilibrium position from a small displacement, and secondly where surfactant is present in the system. The additional entropy production considered in the first case arises from the work done by the interface on the bulk fluid. In the second case surfactant gradients produce additional tangential forces on the interfaces. In these two more complex cases the fluxes and forces identified are somewhat sophisticated: for example the additional fluxes involved are the Fourier components of the displacement of the interface and surfactant concentration from their equilibrium values. Reciprocity was shown to hold in both cases.

The domain of validity of these results and their applicability to results obtained from our simulations is clearly of great interest. Flekk{\o}y and Pride point out that any local nonlinearity (such as bubble breakup) in the flow will invalidate~\ref{eq:recip}. Although the presence of such local nonlinear phenomena may result in an average flow behaviour which obeys a force-flux relationship of the form~(\ref{eq:mpdarcy}), there is no reason that the coefficients $k_{ij}$ should be the same as those for completely linear hydrodynamics. The presence of reciprocity in the results obtained by Olson and Rothman~\cite{bib:olroth2} and in the two dimensional implementation of our model~\cite{bib:pvcjb} to within the simulation accuracy is therefore still a complete mystery from the point of view of linear nonequilibrium thermodynamics.

\subsection{Single phase and binary immiscible flow}

\begin{figure}[htp]
\centering
\subfigure[]{\label{fig:bindarcyforced}\resizebox{0.3\textwidth}{!}{\includegraphics{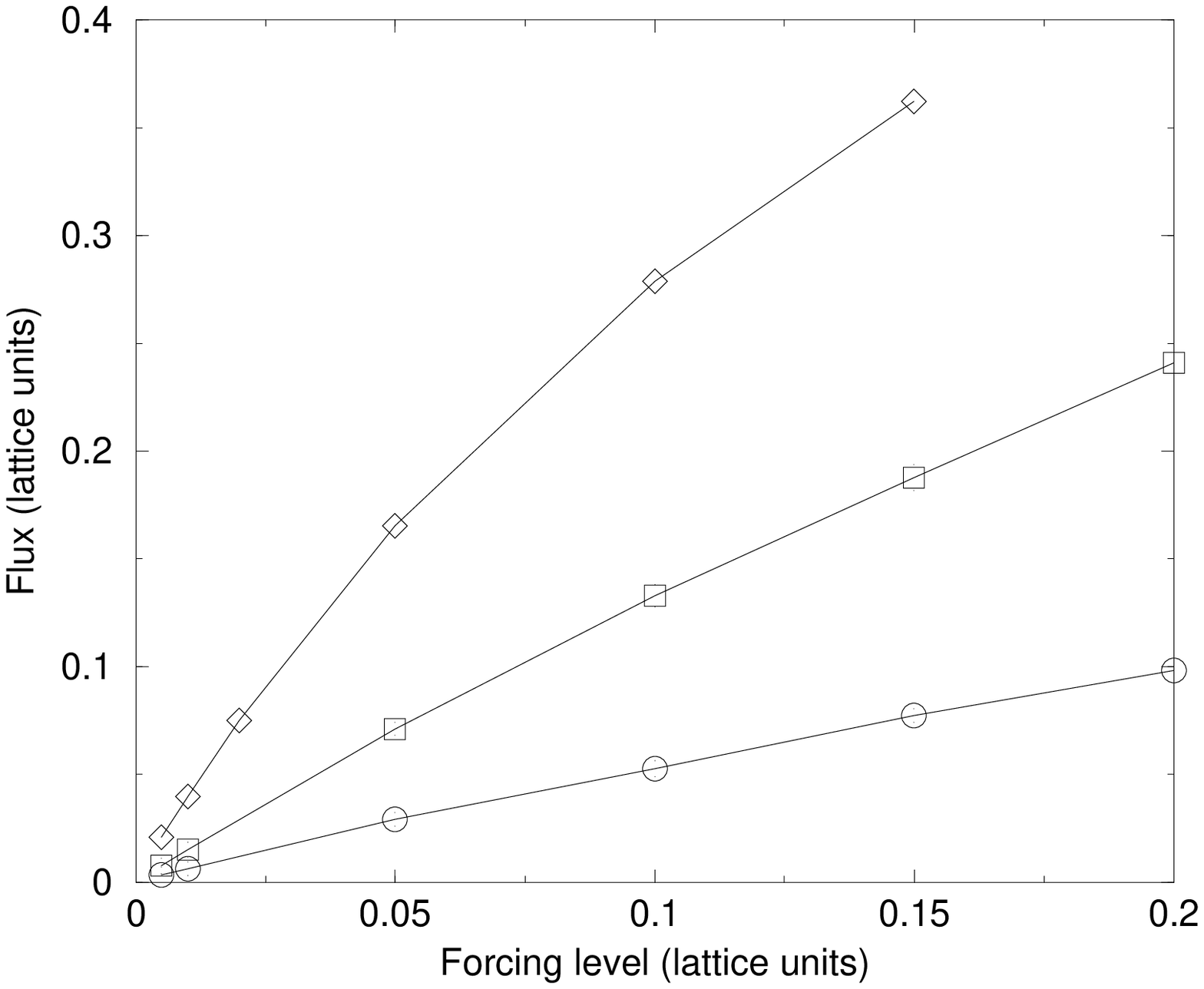}}}
\subfigure[]{\label{fig:bindarcyunforced}\resizebox{0.3\textwidth}{!}{\includegraphics{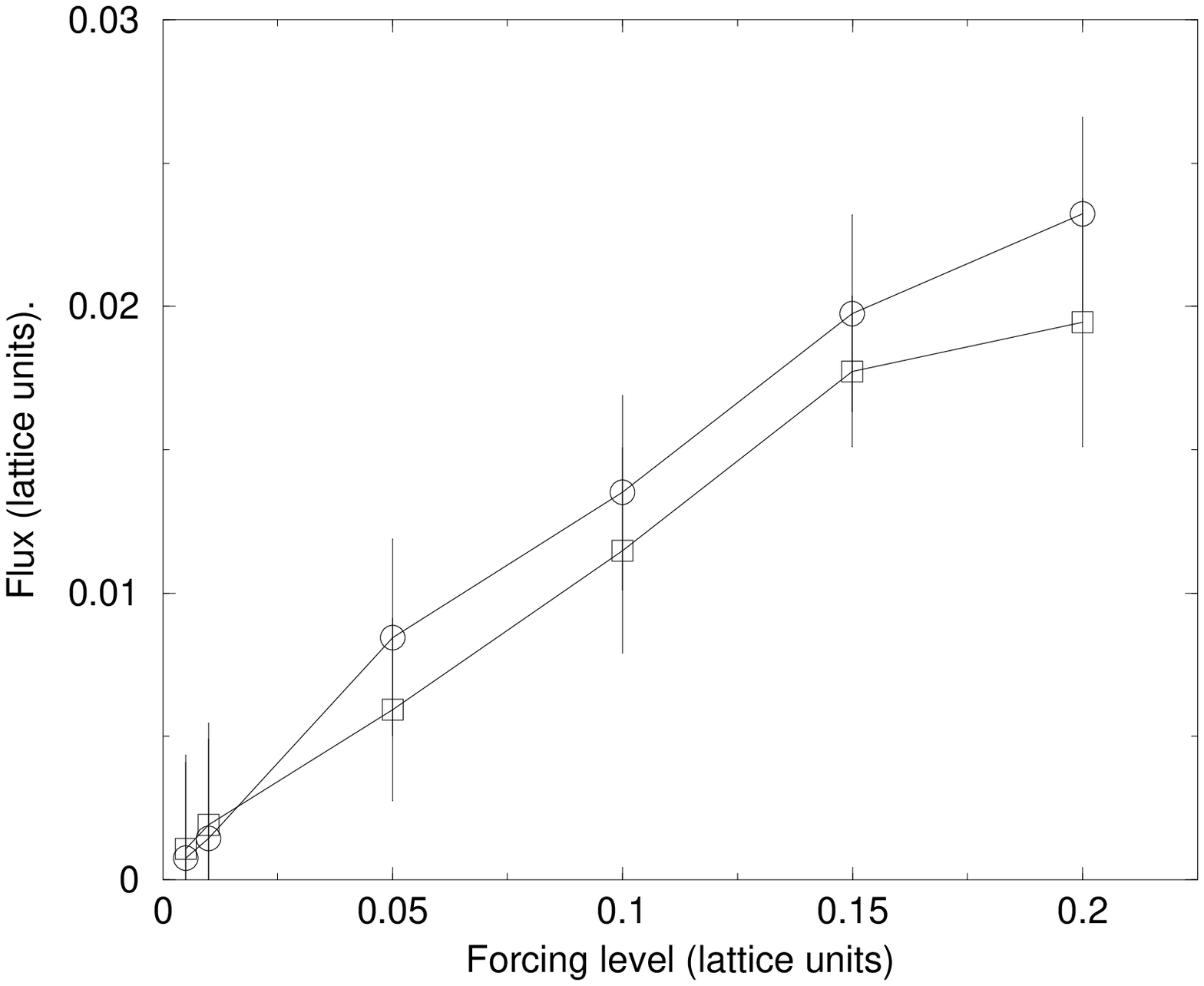}}}
\caption{Verification of Darcy's law for a) Single phase and binary fluids when forced. b) Flux of binary fluids  when unforced. Flux is normalised to show average momentum at each lattice site in the pore space. Diamonds show singl;e phase flux, squares show oil flux and circles show water flux. }\label{fig:binary_ff}
\end{figure}

We performed simulations which address the issues discussed above in the case of binary immiscible flow. Firstly we measured the flux as a function of forcing to investigate the validity of linear relationships such as~(\ref{eq:spdarcy}) and~(\ref{eq:mpdarcy}). We then simulated single phase flow in order to obtain the reference permeability of our medium. Five independent simulations were performed for $2000$ time steps on $16$ processors, for forcing levels $0.005$, $0.03$, $0.05$, $0.1$, $0.15$. The flux was time-averaged after allowing $500$ timesteps for the flow to reach a steady state. The results of these simulations are displayed in Figure~\ref{fig:binary_ff}.

The behaviour of the flux as a function of forcing was then investigated for binary immiscible flow at the same values of forcing as in the single phase case. Two types of simulations were performed: one in which oil was forced, and one in which water was forced. Two independent simulations were performed for $2000$ time steps at each forcing level and for each forced fluid. The flux was time-averaged after allowing $500$ timesteps for the flow to reach a steady state. The results of these simulations are displayed in Figure~\ref{fig:binary_ff}.
 
The squares correspond to the flux of oil when it is forced and
the circles correspond to the flux of water. A linear variation
of the flux versus the force is observed, over the whole range of
forcing levels. The relative permeability coefficient is
smaller in the case of water because of the wettability of the rock. One expects the water phase to flow adjacent to the rock and therefore experience a greater drag force than the oil phase. Visualisation of the flow morphology for our simulations verifies this expectation. As can be seen from figure~\ref{fig:bindarcyunforced} the reciprocity relation $k_{ij} = k_{ji}$ is valid within the statistical uncertainty. 

These results are consistent with those obtained from the two dimensional implementation of our model~\cite{bib:pvcjb2}. The nonlinear deviations from Darcy's law observed at high forcing in two dimensions was not observed in three dimensions. The magnitude of the cross terms in three dimensions is reduced by an order of magnitude. This is consistent with previous three dimensional results, and the magnitude of the cross terms in~\cite{bib:pvcjb2} was believed to be due to the dimensionality of the model. In two dimensions our model was shown to exhibit a capillary threshold at low forcing. The flux of the unforced fluid would only become non-zero at some threshold forcing level. Such effects are not observed in the results presented above, however, the porosity of the medium used here ($0.58$) is close to that of the medium with the largest porosity used in two dimensions ($0.60$). The capillary threshold in two dimensions systematically decreased with increasing porosity. The considerable extra computational cost of performing calculations in three dimensions did not permit a systematic study of the influence of the geometry of the porous medium on Darcy's law behaviour.

\begin{figure}[htp]
\centering
\subfigure[]{\label{fig:binsatdarcy}\resizebox{0.4\textwidth}{!}{\includegraphics{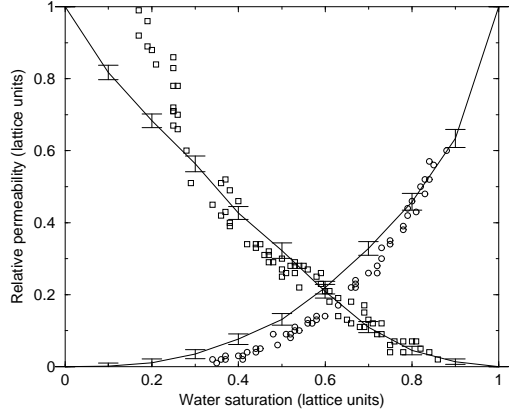}}}
\caption{Variation of the diagonal components of the relative permeability matrix $k_{ij}$ with wetting fluid saturation. Lattice-gas simulation data for a $64^3$ system using the porous medium described above (lines). The medium has porosity $0.58$. Experimental data on $CO_2$/water flow through unconsolidated sand sample of porosity $0.47$ from Wyckoff and Botset (symbols)~\cite{bib:wyckoff}. Squares show gas permeability (non-wetting fluid), circles show water permeability (wetting fluid).}\label{fig:diagdarcy}
\end{figure}

\begin{figure}[htp]
\centering
\resizebox{0.4\textwidth}{!}{\includegraphics{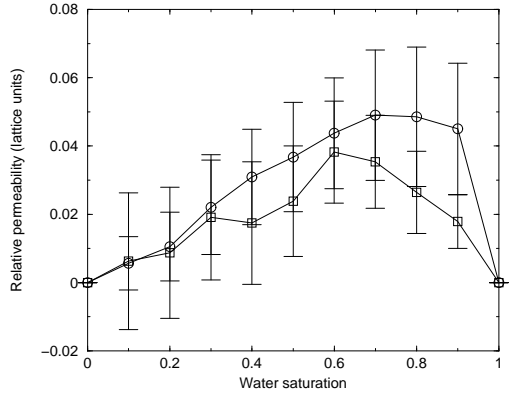}}
\caption{Variation of the off-diagonal components of the relative permeability matrix $k_{ij}$ with wetting fluid saturation. The squares show relative permeability for oil (non-wetting fluid) when water is forced, the circles show relative permeability for water (wetting fluid) when oil is forced.}\label{fig:offdiagdarcy}
\end{figure}

The behaviour of the matrix $k_{ij}$ as a function of the fluid composition is of considerable interest, and enables us to connect the lattice-gas model to previous experiments and network models in which this dependence has been investigated~\cite{bib:goodramak,bib:wyckoff}. In order to perform simulations over a range of compositions within available computational resources the forcing level was kept constant at $0.05$. Simulations were performed at reduced density of water $0.05 \leq \rho_w \leq 0.45$ at intervals of $0.05$. The total reduced density was held constant at $0.5$. Two types simulations were performed, one in which oil was forced, and one in which water was forced. Two independent simulations were performed for $2000$ time steps at each composition and for each forced fluid. The flux was time averaged after allowing $500$ timesteps for the flow to reach a steady state. The results of these simulations are displayed in figure~\ref{fig:binsatdarcy} and \ref{fig:offdiagdarcy}. 

Figure~\ref{fig:diagdarcy} shows the diagonal components of $k_{ij}$ for the experiments conducted by Wyckoff and Botset~\cite{bib:wyckoff}. This data was obtained from flows of water and carbon dioxide in unconsolidated sand packs. The data displayed is for the sand specimen with porosity $0.47$, the closest porosity to that of our virtual Fontainebleau sandstone.

The general features of these graphs can be illuminated considerably by visualisation of the flow morphology in our simulations. For low water (wetting fluid) saturations small bubbles of water exist in a background of oil. These bubbles are predominantly adjacent to the rock as one would expect. Increasing the water saturation increases the coverage of the rock by water, and pores filled with water appear. The morphology of the oil phase is fully connected through the pores until the water saturation reaches $0.8$. For saturation greater than 0.8, there is no connected path of oil through the porous medium.

How do the changes in oil morphology relate to the features seen in figures~\ref{fig:diagdarcy} and~\ref{fig:offdiagdarcy}? First consider the diagonal components of $k_{ij}$. For low water saturations the flux of water is zero below some critical saturation, around $0.2$ for our simulation data and $0.4$ for the comparison experimental data of~\cite{bib:wyckoff}. This can be ascribed to the presence of many small droplets of wetting fluid being adsorbed onto the media and therefore unable to flow. The larger threshold in the experimental case can be interpreted as a consequence of the differences in forcing between the two cases. The water was driven in the experiments by a pressure gradient, and so the net momentum transferred to the fluid will depend on the morphology. In particular, the fluid will not flow until a connected path exists across the medium. This is not the case in our simulations where the forcing reproduces the effects of the wetting fluid being driven by buoyancy forces. 

A similar argument may be applied to the reduction of the non-wetting fluid flux to zero at a saturation of around $0.9$. At this saturation the non-wetting fluid ceases to be connected. It should be pointed out that both these effects are illustrative of a {\it capillary threshold} for the flow. Such a threshold was not observed in our study of the behaviour of flux versus forcing above. However, for the composition used in that study both fluids were continuously connected across the full dimensions of the porous medium. The agreement between the permeability of water in figures~\ref{fig:diagdarcy} and~\ref{fig:offdiagdarcy} is quite good. However, below a water saturation of $0.3$ the non-wetting permeabilities are quite different. This presumably arises because the experimental data is obtained for water and carbon dioxide, i.e. a liquid/gas flow where the fluids have very different hydrodynamic behaviour. Our immiscible fluids are hydrodynamically identical. An extensive study of the dependence of $k_{ij}$ on the viscosity ratio of the two fluids was performed in~\cite{bib:goodramak}, and features such as that reported in figure~\ref{fig:diagdarcy} below a water saturation of $0.3$ were obtained.

The statistical uncertainties on the off-diagonal components of $k_{ij}$ make it impossible to formulate any definitive statements; nevertheless a few comments are appropriate. Firstly we see that, within the statistical uncertainties $k_{ij}$ is a symmetric matrix. As pointed out above this is not a consequence of Onsager reciprocity, and future work should provide a more rigorous test of the symmetry properties of $k_{ij}$ in the non-linear flow regimes accessible to our model. Secondly, the maximum flux of the unforced fluid occurs at a water saturation of $0.8$, that is, at the percolation transition for the unforced fluid. This is consistent with the intuitive picture that viscous coupling is reduced for a flow of bubbles in a background phase.

These results show the same qualitative connection between connectivity of the oil and water phases and the features of the $k_{ij}$ dependence on saturation. The significant exception is the absence of lubrication effects in three dimensions. The oil (non-wetting) flux at low water saturation was significantly larger than that of a pure oil phase in two dimensions~\cite{bib:pvcjb}. No such effect was observed here. 

\subsection{Ternary amphiphilic flow}

In this section the influence of surfactant on the linear flux-force description is investigated. We retain the form of the multiphase Darcy's law presented above. In this case the relative permeability matrix $k_{ij}$ is a $3 \times 3$ matrix which explicitly allows coupling between any pair of species. The influence of surfactant will be described with reference to the questions addressed above, namely a) When is the linear flux-force relationship valid for ternary flow? b) What symmetry properties can be demonstrated for the matrix $k_{ij}$? c) What is the influence of surfactant on the magnitudes of the diagonal and off-diagonal components of $k_{ij}$ as compared with the binary case? In the two dimensional version of the model the significant effect of surfactant was to reduce the capillary threshold by reducing the oil-water interfacial tension. We do not observe a capillary threshold in the binary Darcy's law simulations and so perform simulations to investigate the influence of surfactant on the general features of the binary Darcy law behaviour discussed above.

\begin{figure}[htp]
\centering
\resizebox{0.8\textwidth}{!}{\includegraphics{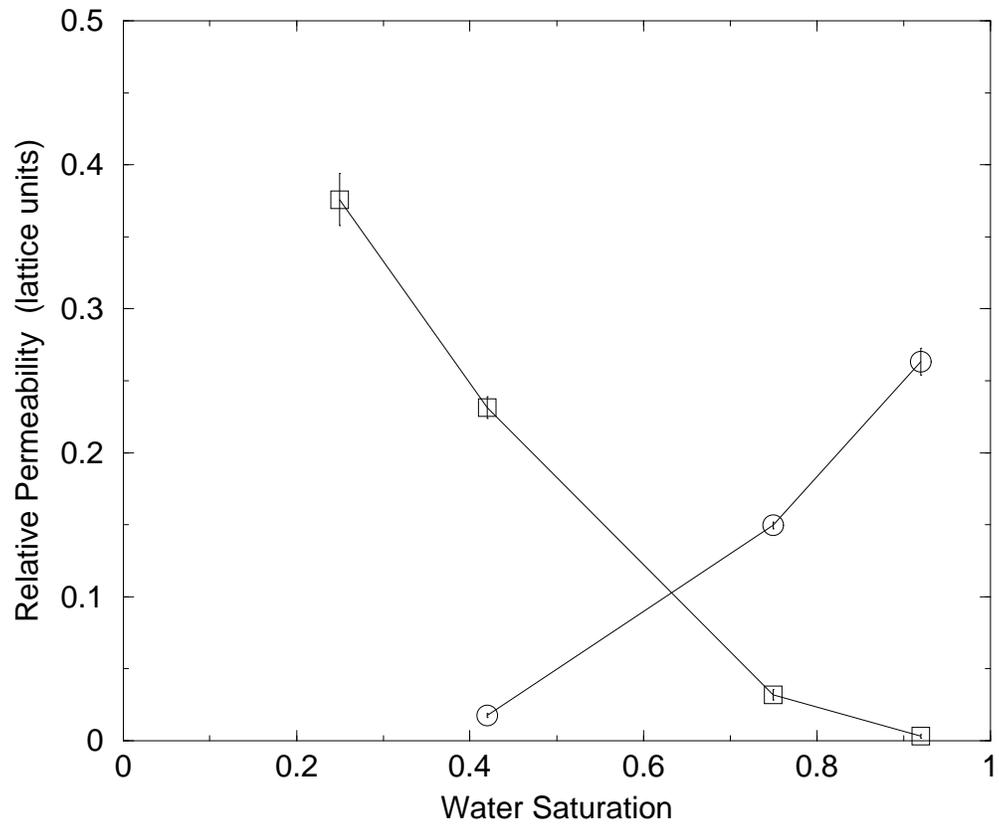}}
\caption{Variation of the $k_{11}$ (squares) and $k_{22}$ (circles) components of the relative permeability matrix $k_{ij}$ with water saturation for ternary amphiphilic steady-state flow.}\label{fig:terndiag}
\end{figure}

We performed simulations for $2000$ time steps for $5$ forcing values. Two types of simulations were performed, one in which oil was forced, and one in which water was forced. Four values of water reduced density were investigated: $0.05$, $0.15$, $0.35$ and $0.45$. The surfactant reduced density and total reduced density were kept constant at $0.1$ and $0.6$, respectively. Fluxes were averaged over for $1500$ time steps after allowing $500$ timesteps for the flow to reach a steady state. Because simulations in which the surfactant is forced were not performed the $k_{13}$,$k_{23}$ and $k_{33}$ components of $k_{ij}$ were not investigated in these simulations. The $k_{11}$ and $k_{22}$ components of $k_{ij}$ are plotted as a function of water saturation in figure~\ref{fig:terndiag}. These diagonal components are considerably reduced as compared with the binary case. 

\begin{figure}[htp]
\centering
\subfigure[]{\label{fig:owternarycouple1}\resizebox{0.4\textwidth}{!}{\includegraphics{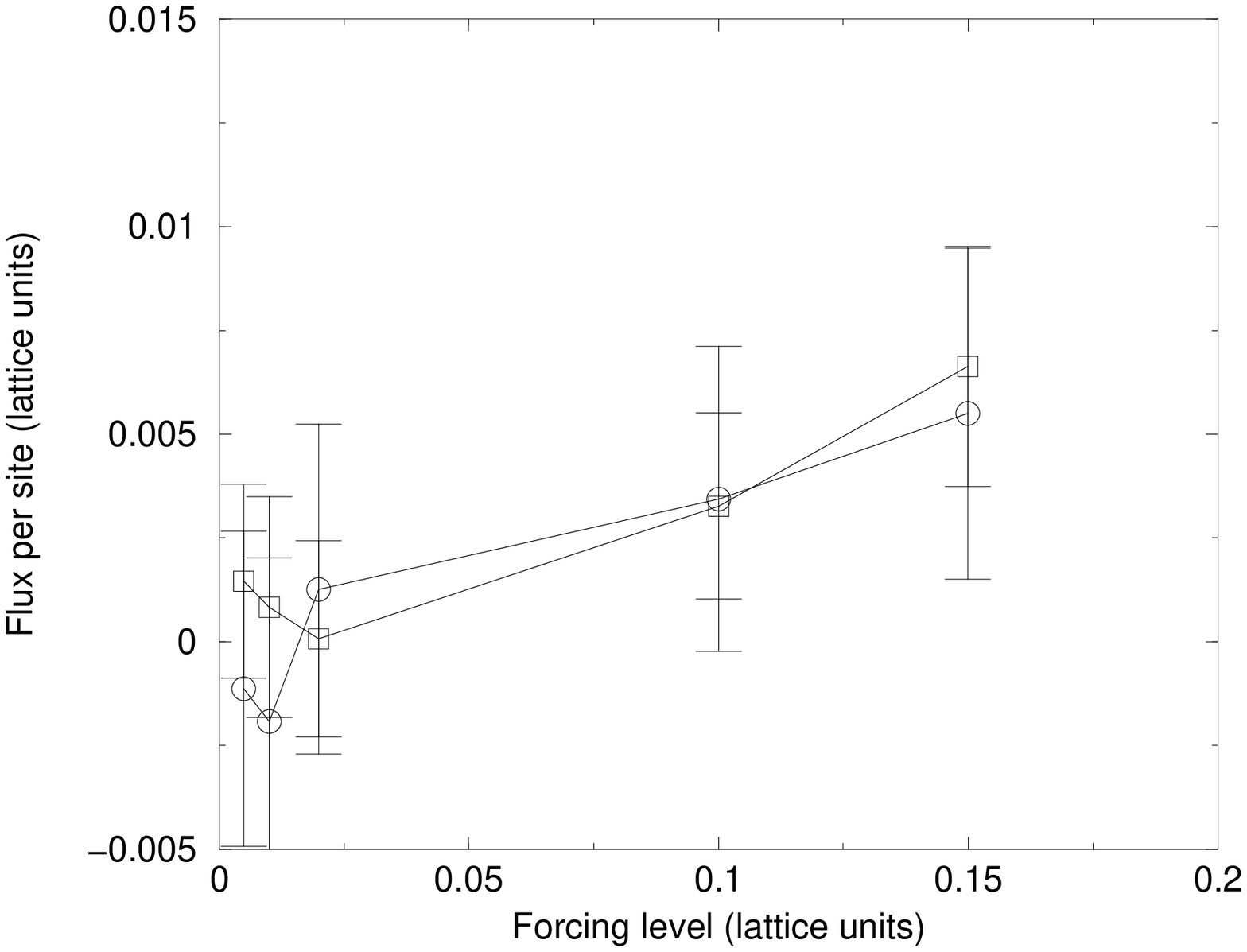}}}
\subfigure[]{\label{fig:owternarycouple2}\resizebox{0.4\textwidth}{!}{\includegraphics{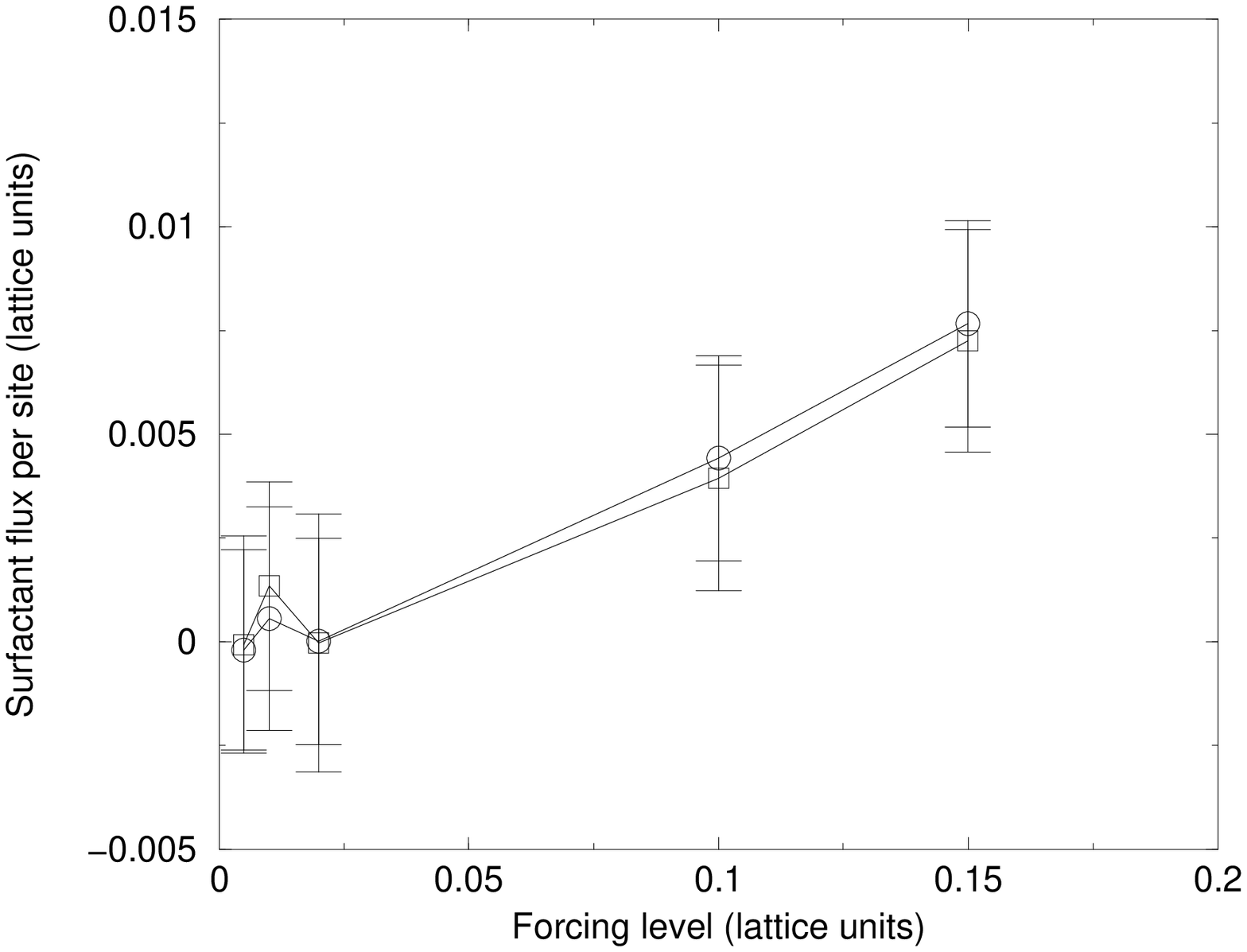}}}
\caption{Fluid-fluid coupling for steady state ternary amphiphilic flow. System size $64^3$.  a) Oil flux (squares) and water flux (circles) when water and oil are forced, respectively. Fluxes measured for a water saturation of $0.7$. b) Surfactant flux when oil is forced (squares), and when water is forced (circles).}\label{fig:owternarycouple}
\end{figure}

The influence of the surfactant on the fluid-fluid coupling may be investigated by examining the behaviour of the off-diagonal components of the relative permeability matrix. For low water saturations the measured fluxes of the unforced fluid are swamped by the statistical errors. A comparison of the flux-forcing relationship for water saturation $0.7$ is shown in figure~\ref{fig:owternarycouple1}. Within the statistical errors this sector of the relative permeability matrix appears to remain symmetric in the presence of surfactant. In figure~\ref{fig:owternarycouple2} the surfactant flux when oil or water are forced is shown, as a function of forcing. This validates the extension of Darcy's linear relationship, eqn~(\ref{eq:mpdarcy}) to the three phase case, at least for ternary amphiphilic flows. A linear response of the surfactant to oil forcing is also seen for water saturation $0.05$ and $0.15$. 

Figure~\ref{fig:owternarycouple2} implies that, within statistical uncertainty, $k_{31}=k_{32}$ at this water saturation. This deserves some consideration. Clearly, any postulated symmetry of $k_{ij}$ does not imply $k_{31}=k_{32}$. The microdynamics of our model is invariant under interchange of oil and water particles. However, because the wettability of the rock as defined above is $-26$, macroscopic parameters such as $k_{ij}$ are not symmetric with respect to interchange of oil and water. If most of the surfactant is adsorbed to interfaces in the flow far from solid-fluid boundaries, one might expect the water-surfactant and oil-surfactant coupling to be symmetric. Without accurate data for a range of water saturations it is of course possible that the agreement displayed in~\ref{fig:owternarycouple2} is coincidental. 

Visualisation of the flow morphology for these simulations confirms that the surfactant is behaving as expected. The preference of surfactant to adsorb at oil-water interfaces leads to an additional effect for the water wetting medium studied here. An oil-solid boundary plays the same role as a water-surfactant interface, and so we expect surfactants to adsorb to the rock in such cases, thereby modifying the wetting properties of the rock. Such behaviour has been observed in our simulations. The surfactant can take on three roles: firstly it can exist in bulk oil or bulk water as (at this concentration) wormlike micelles. When the saturation of one fluid is much greater than the other this is the predominant behaviour. The micelles are carried along in the flow, leading to the water-surfactant and oil-surfactant coupling discussed above. Secondly the surfactants adsorb to oil-water interfaces as expected. Thirdly the surfactant adsorbs to the porous media. For intermediate saturations all the oil flows either in pores coated by a thin layer of surfactant, or if water coats the rock surfaces a layer of surfactant is adsorbed to the oil-water interface. The tendency of the surfactant to enable oil to flow adjacent to the rock, rather than being lubricated by a layer of water, may be one factor contributing to the reduction of the diagonal components of $k_{ij}$ discussed above. 

\section{Imbibition Simulations}

The steady state studies in the previous section attempt to characterise the flow in terms of the simple Darcy's law phenomenological description. Such phenomenological relationships are frequently used as constitutive relationships in coarse-grained continuum models. These models are used to study the non-steady state flows of interest in oil-resevoir flows and ground water pollution remediation. Such an approach could be described as {\it ad hoc}, at best. One advantage of the lattice-gas technique is that the model provides a general description of the flow, valid for both steady and non-steady flows. 

Potentially one of the most commercially rewarding applications of surfactants is their use in enhanced oil recovery. In oil field flows, oil is displaced and carried to the surface by water. Two limiting factors exist for this technique. Firstly, the ratio of oil to water must be high enough for the process to be commercially viable. Secondly, the total amount of oil extractable is limited. As one might expect, the immiscibility of oil and water eventually confounds efforts to solubilise one into the other. These limitations mean that most commercially developed oil fields have between $40 \%$ and $60 \%$ of the available oil remaining after primary and secondary recovery operations. Addition of surfactant to the extraction fluid should increase the solubilisation of oil into the water, thereby increasing well productivity. It should be noted that the amounts of surfactant needed and their associated cost mean such techniques will not find application until the world oil price has increased substantially.

We perform simulations intended to reproduce such oil field extraction flows. The porous media is initially saturated with oil at a reduced density of $0.5$, and a water surfactant mixture is forced into one end of the media. The subsequent flow morphology and time history of oil concentration remaining in the media are then studied. The wettability of the porous media is kept constant $-26$, i.e. completely water wetting. Invasion into such a porous medium is referred to as {\it imbibition}, from the medium's natural tendency to {\it imbibe} the wetting fluid even in the absence of forcing. These simulations extend work performed in two dimensions by Fowler and Coveney and Maillet and Coveney~\cite{bib:pvcjb,bib:pvcjb2}.

We vary two parameters in the invasion study, the fluid
forcing level and the fraction of surfactant in the invasive
fluid. The range of input forcing is limited by two constraints. The
forcing must be low enough that the underlying lattice-gas dynamics
still reproduces Navier-Stokes behaviour, and must be high enough that
the simulation reaches the percolation point within simulation lengths attainable with current resources.

\begin{figure}[htp]
\begin{center}
\resizebox{0.8\textwidth}{!}{\includegraphics{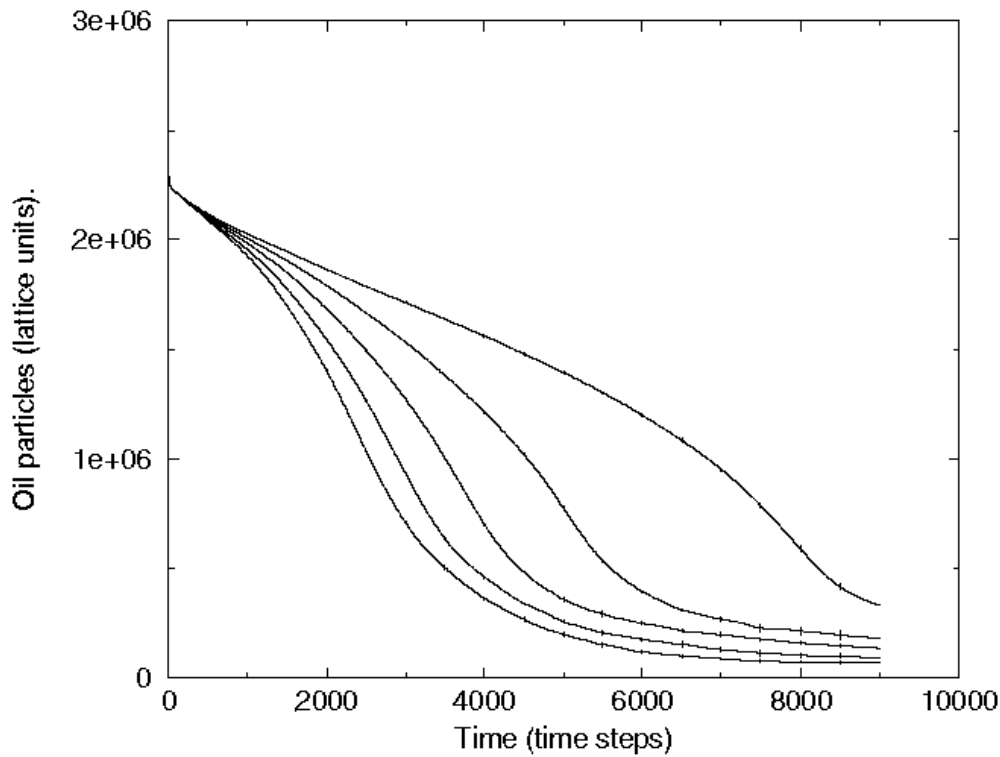}}
\caption{Number of oil particles in system as a function of time for
  various forcing levels in a binary system. The forcing level is from
  right to left $0.01, 0.02, 0.03, 0.04, 0.05$ (lattice units). Error bars show the standard deviation on the mean of five independent simulations.\label{fig:binoilforcing}}
\end{center}
\end{figure}

We first study the extraction of oil by an invasive
fluid comprising pure water. Five independent simulations were performed for each forcing level at forcing levels $0.01$, $0.02$, $0.03$, $0.04$, and $0.05$. The temporal evolution of the oil saturation is shown in figure~\ref{fig:binoilforcing}. There are three production regimes observed here. By visualising the densities of oil and water we may identify these as follows: an initial slow regime prior to water percolation, a regime subsequent to water percolation and prior to oil depercolation in which the establishment of high flow rate channels causes rapid oil production, and a third regime after the end of oil percolation in which the oil saturation approaches its asymptotic value. 

In the two dimensional version of our model the first and last of the extraction regimes described above were observed. This may be a dimensional effect, as the second regime in which oil is extracted rapidly by water domains connected across the medium requires both oil and water domains which extend across the medium. Such a bicontinuous morphology of oil and water in the porous media can only occur in three dimensions. 

\begin{figure}[htp]
\begin{center}
\resizebox{0.8\textwidth}{!}{\includegraphics{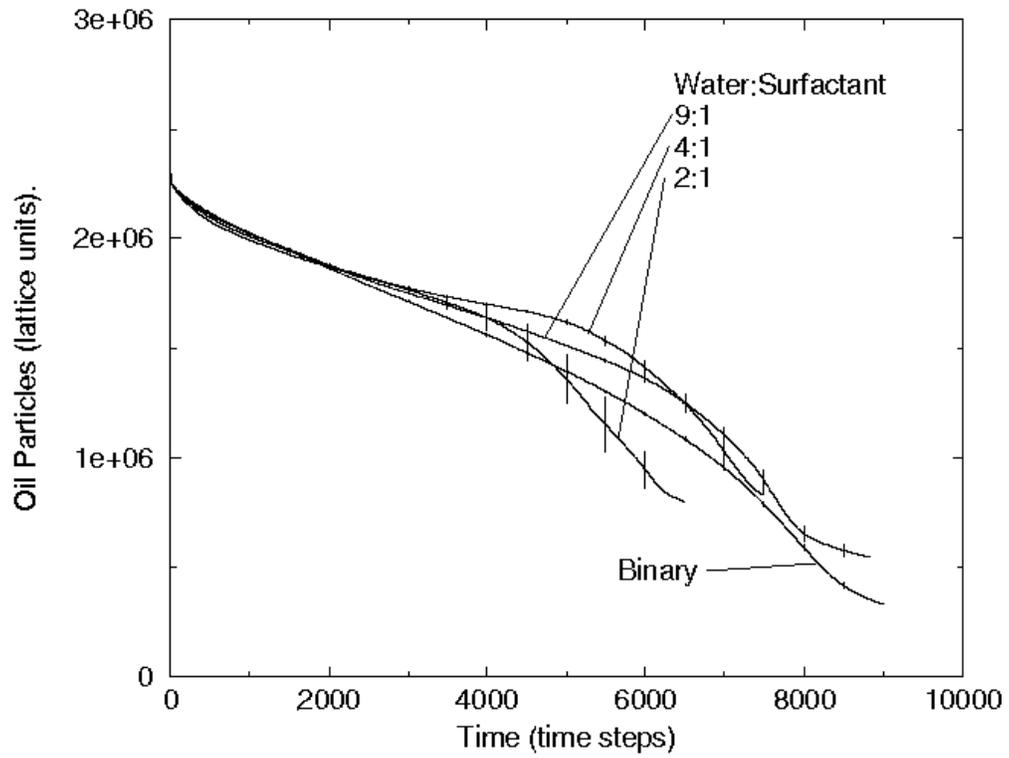}}
\caption{Number of oil particles in system as a
  function of time for various invasive fluid compositions in a ternary system. Error bars show the standard deviation on the mean of five independent simulations.\label{fig:oil_density_vs_surf_0.05}}
\end{center}
\end{figure}

As pointed out by Maillet and Coveney, the third regime here is of most interest in the context of enhanced oil extraction. In order to investigate the influence of surfactant on the flow morphology and pre-percolation behaviour $5$ independent simulations were performed for each value of forcing and invasive composition, for the $5$ forcing levels used in the binary case, and for invasive fluid water:surfactant ratios of $4:1$ and $2:1$. These are ratios which lead to equilibrium phases consisting of wormlike micelles, droplet phases and sponge microemulsions respectively. It is expected that the flow conditions and porous media will cause significant deviations
from the equilibrium morphology. The time evolution of oil saturation as a function of surfactant concentration is shown in figure~\ref{fig:oil_density_vs_surf_0.05}.

This graph has a number of notable features. Firstly, the production
rate prior to water percolation is reduced by the presence of
surfactant. We attribute this to the reduction of surface tension
caused by the surfactant, which leads to a more convoluted interface,
rather than the `piston-like' displacement we observed in the binary
invasion case. However, the surfactant does systematically reduce the time 
required for water percolation as a consequence of the more convoluted
interface. The reduction in oil production relative to the binary system at early times in the ternary system was also reported for the two dimensional version of our model by Coveney {\it et al.}. Only by performing simulations which reached longer time scales could Maillet and Coveney assess the impact of surfactant on the residual oil saturation~\cite{bib:pvcjb2}.

\begin{figure}[htp]
\begin{center}
\resizebox{0.8\textwidth}{!}{\includegraphics{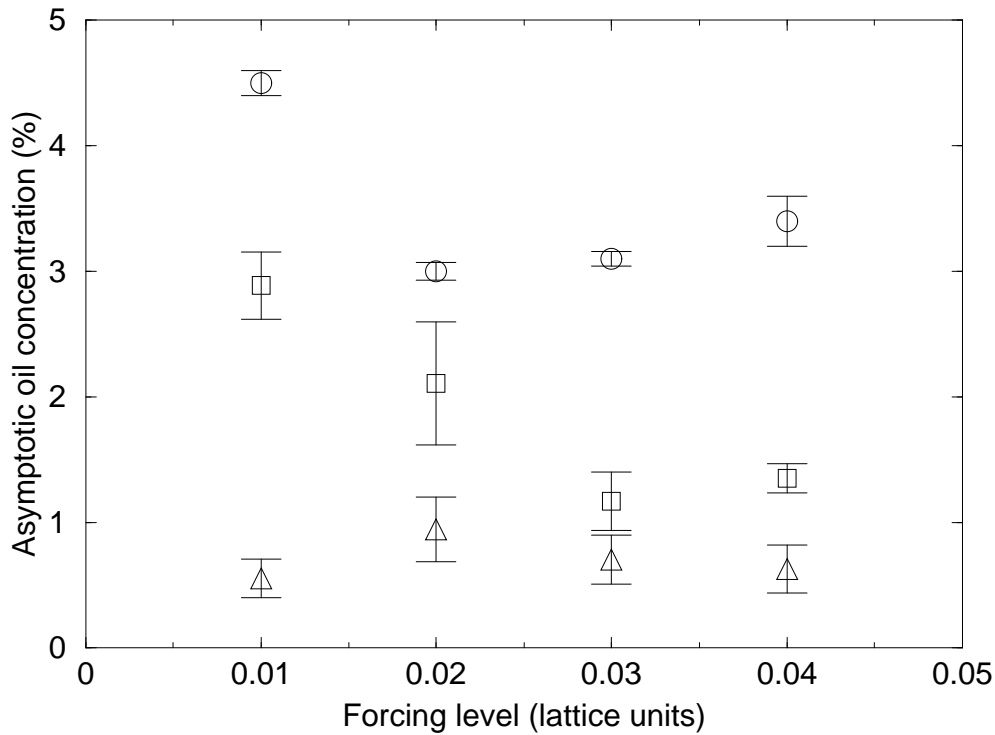}}
\caption{Residual oil saturation (\%) for binary (circles), water:surfactant ratio $4:1$ (squares) and $2:1$ (triangles). \label{fig:asymptotic_log}}
\end{center}
\end{figure}

In order to assess the impact of surfactant on the asymptotic oil saturation further simulations were performed for forcings $0.01, 0.02, 0.03$ and $0.04$. Invasive fluid water-surfactant ratios of $4:1$ and $2:1$ and binary invasion without surfactant were simulated for $20000$ timesteps. The oil saturation reaches a steady state value after $15000$ timesteps. The residual saturation was averaged over the last $5000$ timesteps and a statistical error derived from the standard deviation on the mean. The results are plotted in figure~\ref{fig:asymptotic_log}.

\begin{figure}[htp]
\centering
\subfigure[]{\label{fig:asymp_vis1}\resizebox{0.25\textwidth}{!}{\includegraphics{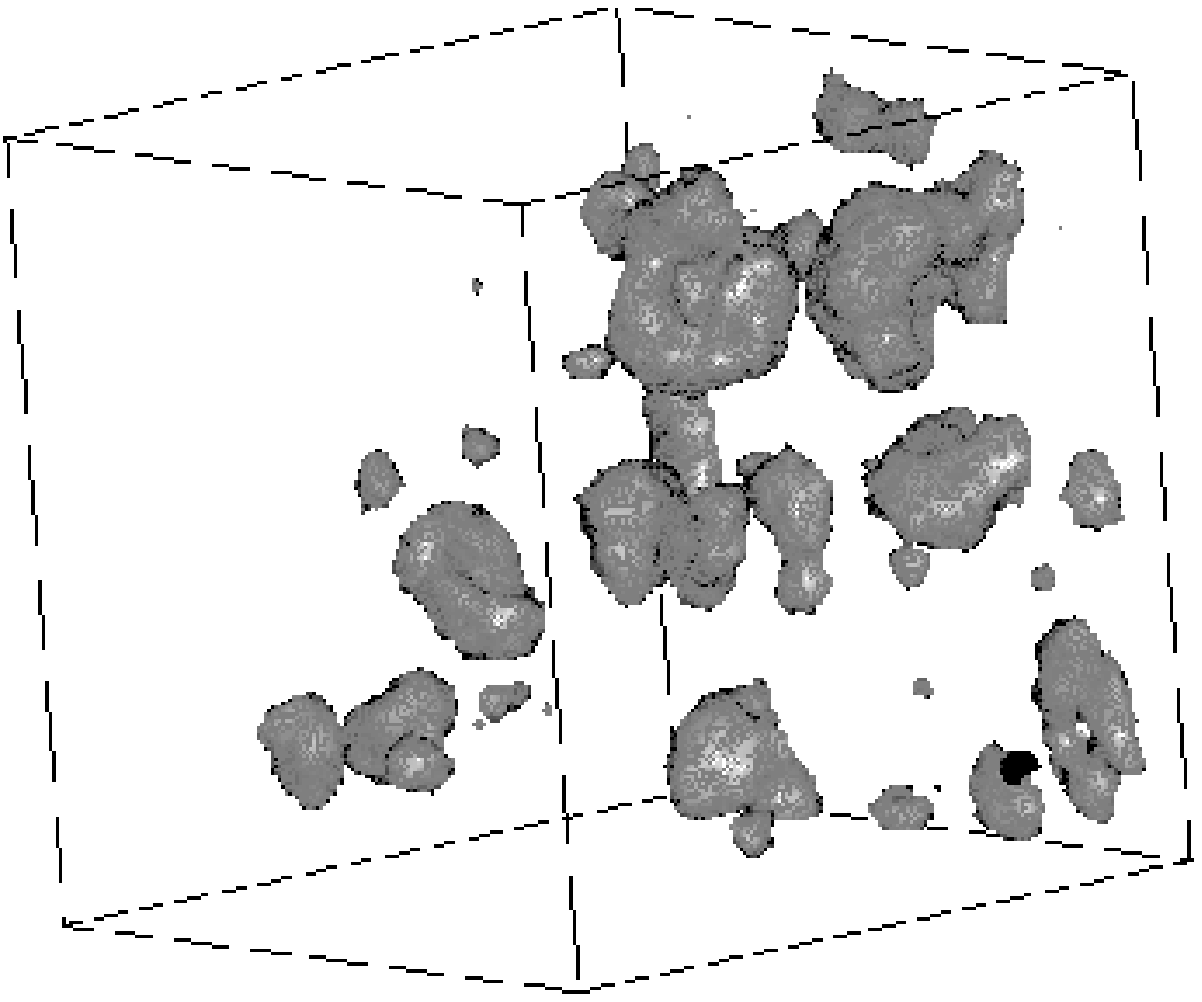}}}
\subfigure[]{\label{fig:asymp_vis2} \resizebox{0.25\textwidth}{!}{\includegraphics{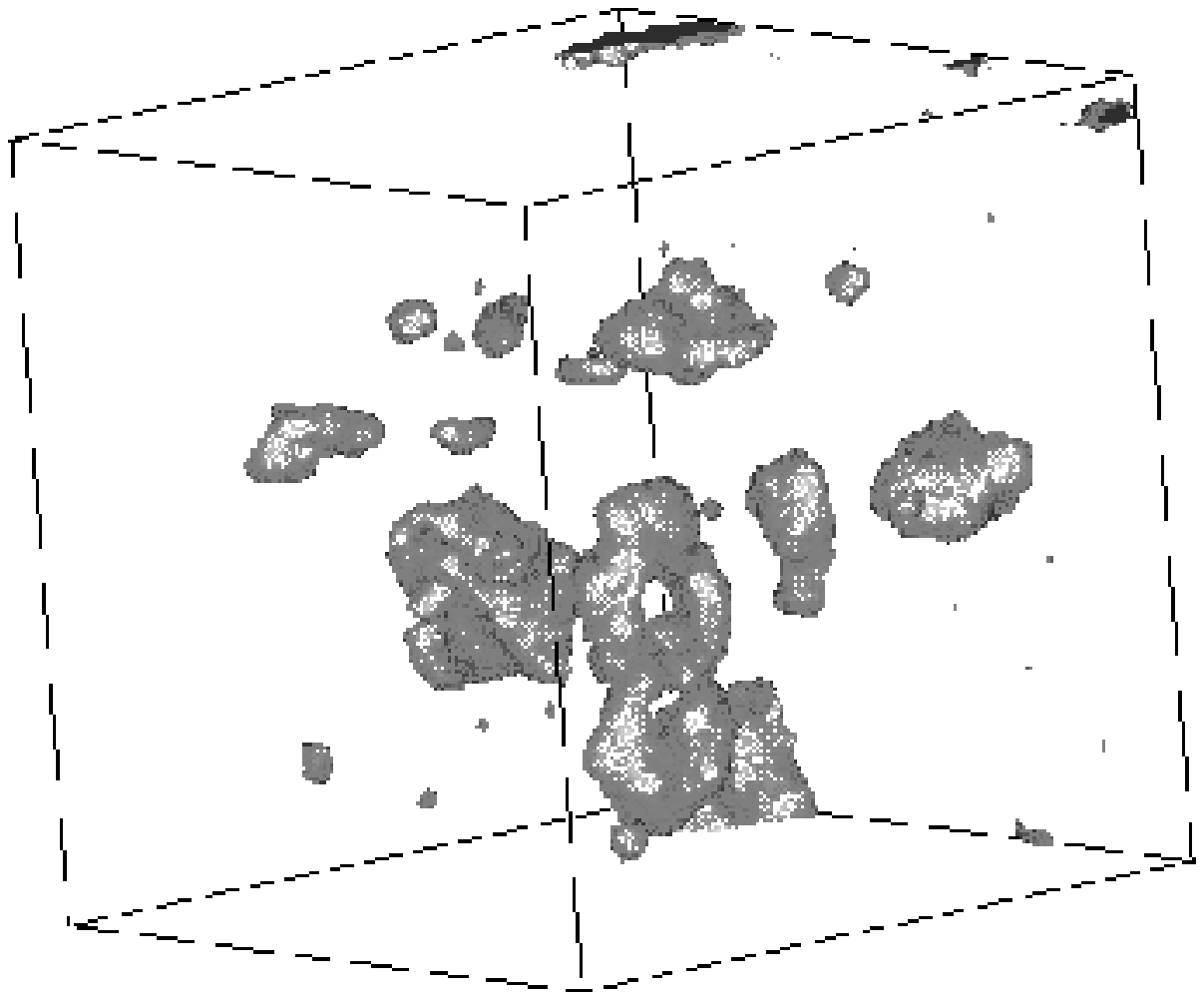}}}
\subfigure[]{\label{fig:asymp_vis3} \resizebox{0.25\textwidth}{!}{\includegraphics{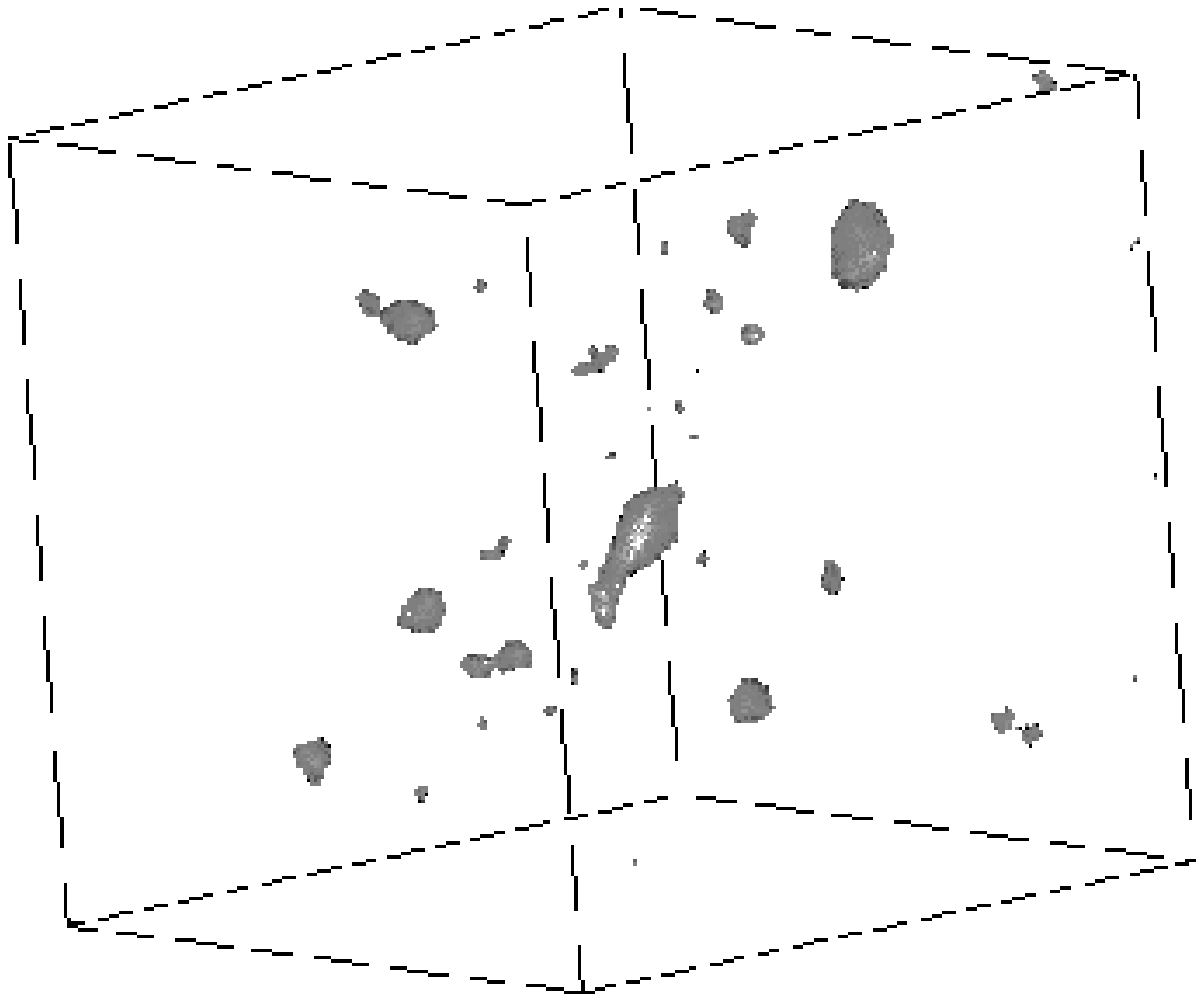}}}
\caption{Isosurface of residual oil at timestep $20000$. a) No surfactant. b) Water-surfactant ratio $4:1$. c) Water-surfactant ratio $2:1$. Isosurface shows oil concentration at a value of $5$ particles per site. Forcing level $0.01$. Porous media is not displayed for clarity.}\label{fig:asymp_vis}
\end{figure}

Figure~\ref{fig:asymptotic_log} bears out our supposition that addition of surfactant to the extraction fluid should have a significant impact on the residual oil saturation in the media. In fact, the residual oil saturation is reduced by a factor of five between the binary case and the case with the highest water-surfactant ratio in the extraction fluid. This is illustrated dramatically by visualising the oil saturation in the medium at time step $20000$ for the three types of invasive fluid, as shown in figure~\ref{fig:asymp_vis}.

These results are in complete agreement with those obtained in two dimensions by Maillet and Coveney~\cite{bib:pvcjb2}. In the two dimensional case the enhancement of oil recovery was shown to be due to the emulsification of the residual oil into the extraction fluid as surfactant coated droplets. In the next section we visualise the behaviour of the surfactant and compute structural characteristics of the interface to investigate the mechanisms involved in recovery enhancement in three dimensions.

\section{Interfacial Morphology}

To better understand the role of surfactant in enhanced recovery we analyse the behaviour of the oil-water interface prior to invasive fluid percolation as a function of surfactant concentration. Visualisation of the interface shows that it is much more convoluted in the ternary case. Maillet and Coveney computed the fractal dimension of such interfaces in the two dimensional version of our model. This fractal dimension was largely determined by the fractal dimension of the porous medium itself. To avoid the additional complexity of such calculations in three dimensions we compute the interfacial distribution function:
\begin{equation}
f(z) = \frac{\sum_{xy} I_i({\bf x})}{\sum_{xyz} I_i({\bf x})},
\end{equation}
where $I_i({\bf x})$ takes the value one if the site at ${\bf x}$ is at the interface and takes the value zero otherwise. Interfacial sites are defined as those at which the colour charge is zero. This function is useful as it allows us to define precisely the interfacial position:
\begin{equation}
{\bar z} = \sum_z f(z) z.
\end{equation}

\begin{figure}[htp]
\begin{center}
\resizebox{0.8\textwidth}{!}{\includegraphics{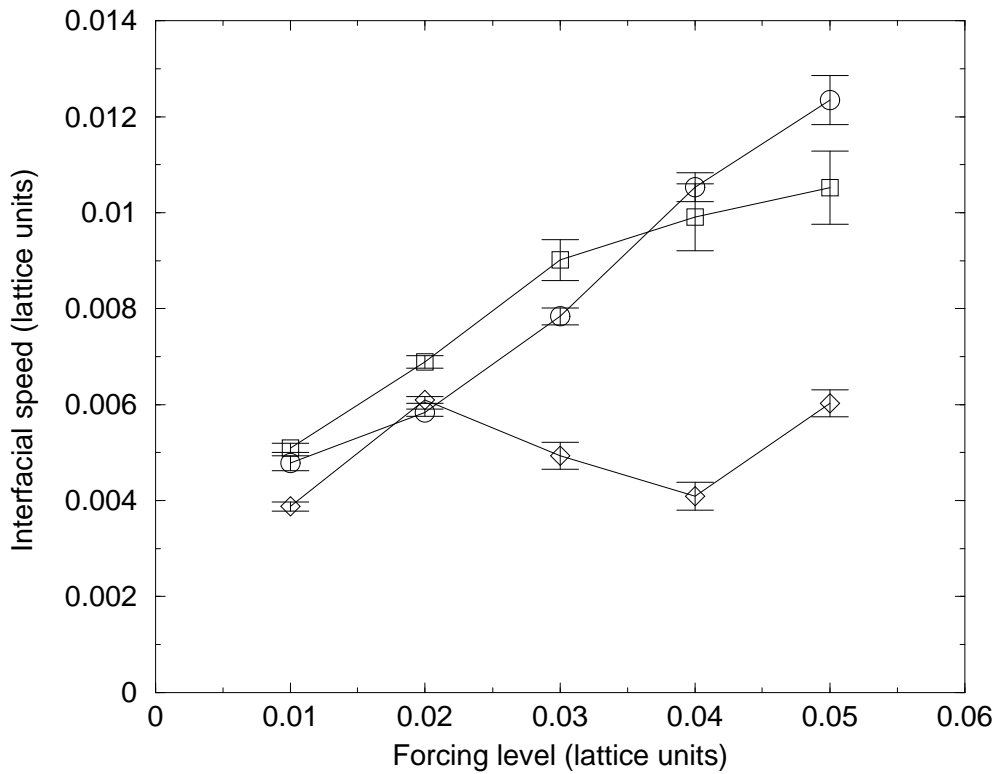}}
\caption{ Interfacial speed $\dot{\bar{z}}$ for binary (circles), water:surfactant ratio $4:1$ (squares) and $2:1$ (diamonds).\label{fig:interface_speed}}
\end{center}
\end{figure}

\begin{figure}[htp]
\begin{center}
\resizebox{0.8\textwidth}{0.8\textwidth}{\includegraphics{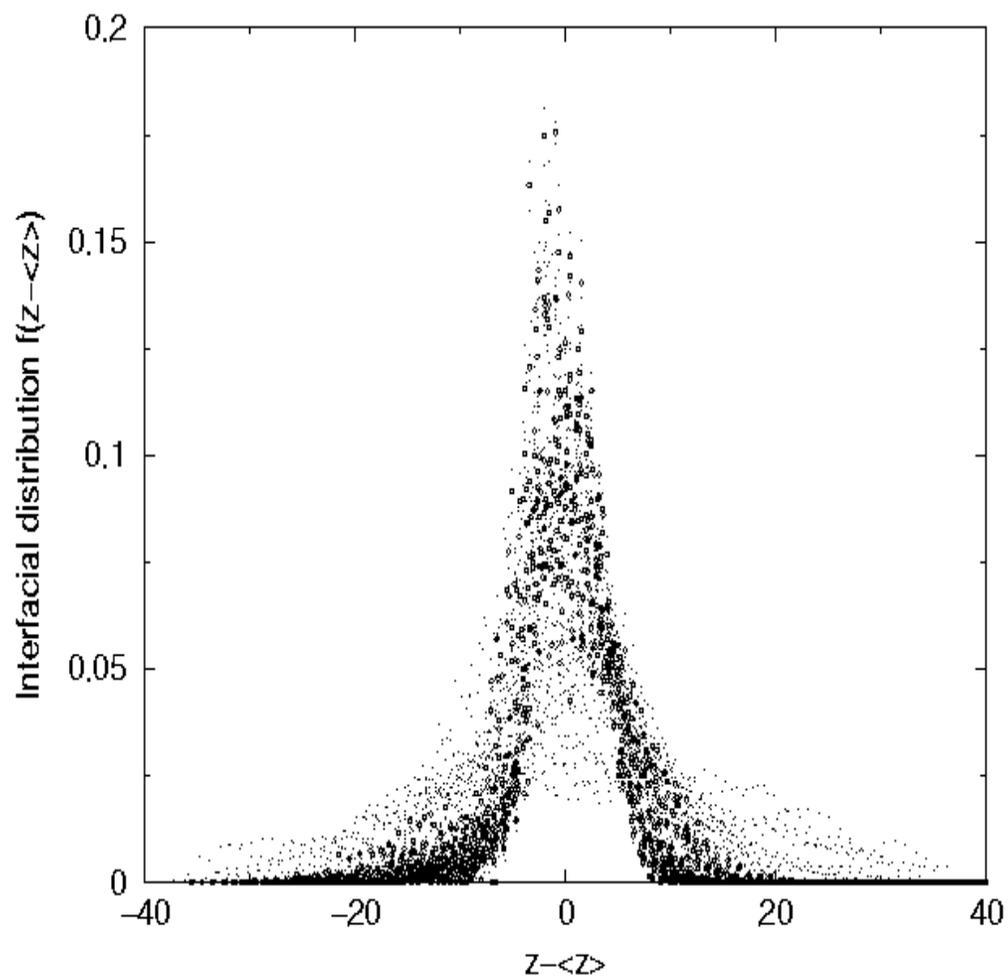}}
\caption{Scaled interfacial distributions for binary invasion (squares) and invasive water:surfactant ratio $2:1$ (points).\label{fig:interfacial_scaled}}
\end{center}
\end{figure}

Prior to percolation the interface moves with constant speed through the system in both binary and ternary simulations. The interfacial speed, defined as $\dot{\bar{z}}$, is plotted in figure~\ref{fig:interface_speed}. The speed for low or intermediate surfactant concentration is proportional to the forcing level as expected. This result is equivalent to the linear dependence of early time extraction rate on forcing level observed in the two dimensional implementation of our model. For high surfactant concentration the interfacial speed appears to be independent of forcing, a somewhat suprising result. From direct visualisation of the fluid interfaces it is clear that the picture of a well-defined interface moving coherently through the porous media is inaccurate for high surfactant concentration. One anticipates that the surfactant will cause significant growth in the width of the interface as a result of emulsification processes. In figure~\ref{fig:interfacial_scaled} we show the interfacial distribution function as a function of position relative to $\bar z$ for binary fluid invasion and for water-surfactant ration of $1:2$. All values of forcing are plotted.

In the binary case significant growth of the interface does not occur over the timescale of the simulation. The addition of surfactant clearly causes the interface to grow significantly. The independence of the interfacial speed and the forcing level arises because emulsification processes dominate prior to water percolation.

\section{Conclusions}

The three dimensional lattice gas model for the
non-equilibrium dynamics of amphiphilic fluids has been extended to simulate
multiphase flow in porous media. The description of macroscopic
behaviour of binary immiscible steady state flows by an
extension of Darcy's law with an explicit viscous coupling between
species has been verified. The extension of this Darcy's law description to ternary amphiphilic fluids has also been investigated

In the binary case the dependence of the relative permeabilities for the forced
fluids bear a strong similarity to results obtained from network models and flows in sands~\cite{bib:goodramak,bib:wyckoff}. The cross coefficients satisfy a reciprocal relationship $k_{12}=k_{21}$, within the statistical uncertainty of the lattice-gas model. As pointed out by Olson and Rothman in~\cite{bib:olroth2}, and~\cite{bib:flekkoypride,bib:prideflekkoy} this is not a consequence of Onsager reciprocity, but is nevertheless now a reasonably well established relation without theoretical foundation. The cross-coefficients are much smaller than those observed in the two-dimensional realisation of our model, as expected.

In the ternary case the linear flux-forcing relationship has been verified for intermediate wetting fluid saturations. The relative permeabilities are reduced by the presence of surfactant. The presence of an additional species increases the difficulty of studying these systems. It should be noted that a systematic study of Darcy's law behaviour as a function of surface tension is possible in the model by varying the inverse temperature-like parameter $\beta$. Such a study could elucidate which aspects of the ternary behaviour are due to the modification of surface tension by surfactant and which are due to other factors, such as modification of the wetting properties of the media.

Invasion simulations performed with both binary and ternary
amphiphilic fluids show three regimes for production of the defending fluid (oil). Prior to invasive
fluid percolation the oil is produced slowly, after percolation and
prior to the end of oil percolation rapid production occurs,
followed by a third regime in which the oil saturation approaches
an asymptotic residual value. Addition of surfactant to the invasive fluid reduces the asymptotic residual oil saturation by a factor of five. The pre-percolation interfacial behaviour is also strongly affected by the presence of surfactant. The interfacial behaviour changes from a `piston-like' movement of an interfacial region of fixed width to a motion dominated by the growth of the interfacial region. The contributions to the behaviour of capillary fingering and other surfactant-induced emulsification processes could be elucidated by a systematic study of invasive flows as a function of surface tension. Such a study would be possible within our model, again by varying the inverse temperature-like parameter $\beta$.

\section*{Acknowledgements}

We are indebted to numerous people and organisations for their support
of and contributions to this work. They include Bruce Boghosian, Keir Novik, Nelido Gonzalez, Jonathan Chin, and Julia Yeomans. Simulations were performed on the Cray T3E and Origin$2000$ at the Manchester CSAR service, on the Onyx $2$ at the Centre for Computational Science, Queen Mary, University of London, and on the $O2000$ at the Center for Computational Science, Boston University. Resources for the T3E were allocated under EPSRC grant number GR/M56234 and resources for the O$2000$ were allocated by special arrangement with CSA. PJL
would like to thank EPSRC and Schlumberger Cambridge Research for
funding his CASE studentship award. The authors would also like to thank the European Science Foundation SIMU programme for funding their collaboration.

%\bibliographystyle{prsty}
%\bibliography{my}

\end{document}